# INCIDENCE OF LOW DENSITY METEOROIDS OF THE POLONNARUWA-TYPE

N. C. Wickramasinghe*[1], J. Wallis[2], D.H. Wallis[1], M.K. Wallis[1], N. Miyake[1], S.G. Coulson[1], Carl H. Gibson[4], J.T. Wickramasinghe[1], A. Samaranayake[3], K. Wickramarathne[3] and Richard B. Hoover[1]

[1]Buckingham Centre for Astrobiology, University of Buckingham, Buckingham, UK

[2]School of Mathematics, Cardiff University, Cardiff, UK

[3]Medical Research Institute, Colombo, Sri Lanka

[4]University of California San Diego, La Jolla, CA, 92093-0411, USA

## ABSTRACT

The ingress of micrometeorites of cometary origin with densities below ~ 1 g cm$^{-3}$ into the Earth could average at least 5 tonne per day.  Although much of this is burnt upon entry through the atmosphere as meteors, a non-trivial fraction (~10%) which have sizes of ~ 1 m could end up in the form of Polonnaruwa-type meteorites that have mostly gone unnoticed.

**Corresponding author**: *Professor N.C. Wickramasinghe, Buckingham Centre for Astrobiology, University of Buckingham, Buckingham, UK:  email – ncwick@gmail.com

**Introduction**

There is no doubt that the vast majority of meteoroids that enter the Earth are derived from comets and asteroids. The cometary component consists mainly of ~1 – 10 μm - radius dust from normal outgassing that forms cometary meteor streams, as well as larger fragments that result from sporadic break-up as observed recently in several comets.

Fig.1 gives an example of such a break-up event for comet 73P Schwassmann-Wachmann3 in which sub-fragments are seen to be separated from the main comet nucleus. Meteoroids entering Earth would also include a component from the break-up of extinct comets that have lost most of their volatiles and are the probable parent bodies of carbonaceous chondrites such as the Murchison meteorite.

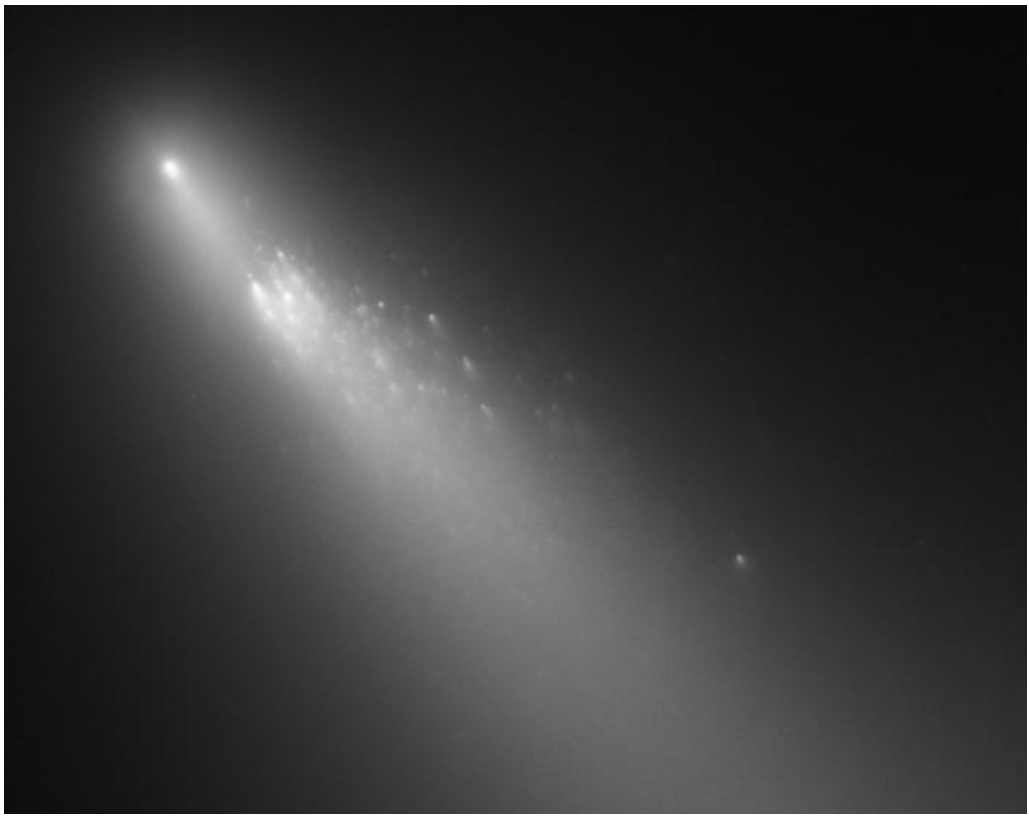

Fig. 1 73P Schwassmann-Wachmann3, fragmenting in 2006 (Courtesy NASA, ESA)

Dust grains with radii upto 10 μm are released from comets near perihelion to form meteor streams that eventually spiral into the sun on timescales in the range ~ 20,000 – 200,000 years (Flynn, 1989; Jenniskens, 1998).

Table 1 shows a compilation displaying the space exposure times (meteor stream lifetimes) and relative velocities of entry into the Earth for 10 μm radius dust with density of 1 gm cm$^{-3}$ adapting data from Flynn (1989).

Table 1: Space exposure age and relative velocity at Earth entry for 10μm radius, 1 g/cc density particles

| Parent comet orbit | Space exposure (yr) | Geocentric velocity (km/s) | Atmospheric entry velocity (km/s) |
|---|---|---|---|
| **Comets (perihelion <1.2AU)** | | | |
| Giacobini-Zinner | 32,000 | 19 | 32 |
| D'Arrest | 40,000 | 13 | 17 |
| Halley | 24,000 | 64 | 65 |
| Swift-Tuttle | 110,000 | 51 | 52 |
| 2P/Encke | 30,000 | 15 | 17 |
| **Comets (perihelion<1.2AU)** | | | |
| Temple 2 | 35,000 | 9 | 14.3 |
| Kopff | 63,000 | 6 | 12.6 |
| Temple 1 | 50,000 | 8 | 13.7 |
| **Comets (near circular)** | | | |
| P/Schwassman-Wachmann 1 | | 5 | 12 |

The case of comet 2P/Encke and the associated Taurid complex is worthy of special note in the present context. This stream of debris that gives rise to regular Taurid meteor showers is thought to comprise of remnants of a giant comet that disintegrated some 20,000 to 30,000 years ago (Ascher and Clube, 1993). The resulting fragments ranging from 10μm-radius dust to fragments possibly larger than the Tunguska bolide are distributed in the stream, being released both by normal cometary activity and occasionally by tidal interactions with planets (Pittich and Klačka, 1993). The Earth takes several weeks to pass through this rather wide stream and thus results in extended periods of meteor activity.

In addition to the distributed dust that is subject to Poynting-Robertson forces and have orbits that evolve independently of the comet, the Taurid stream might also include an "inner" dense region comprised of fragments that remain concentrated near the orbit of the stream's parent object. Ascher and Clube (1993) have argued that the orbits of meteoroids in this dense core have been subject to perturbations over thousands of years, and result in periodic episodes of intense bombardment of the Earth.

**Rate of ingress of meteoroids**

Whenever the Earth encounters a meteor stream the ingress of particles depends on their relative orbital configurations as well as on particle size. If $R_E$ is the radius of the Earth, $v_e = 11.2$ km/s is the escape speed, and $v_p$ is the geocentric velocity of the particle, the accretion cross section is

$$A = \pi R_E^2 [1 + v_e^2 / v_p^2] \tag{1}$$

From Table 1 we can see that the lowest value of geocentric velocity would lead to greater capture cross-sections by a factor of ~ 2. The average rate of meteoroid ingress from all sources has been estimated from various data to be in the range ~ 50 to 300 tonnes per day (Plane, 2012). The upper limit of this range is derived from studies of zodiacal light and from space probes, whilst the lower limit is from radar and lidar observations of meteors in the middle atmosphere. The higher value, which is also consistent with the measured accumulation rates of cosmic elements in polar ice cores and in deep-sea sediments, is probably the better estimate of the total meteoroid input to the Earth. In this article, however, we shall adopt the lower limit of 50 tonnes per day as a *conservative* estimate of the mass flux that reach the ground.

The majority of meteoroids intercepted by the Earth have dimensions of 50-100μm and masses of ~ 10μg, and have their most likely origin in comets. These include short period, Jupiter-family comets for the regular meteor showers, and also contributions from Halley-type and longer period comets. Fig. 2 shows the relative mass influx of meteoroids from data compiled by Flynn (2002). From this figure we can estimate that the observed mass flux of $10^2$-$10^6$g meteoroids is ~ 1/100 of the flux of smaller 10μg IDP's. Adopting the total input to be 50 tonnes per day, we have ~ 0.5 tonne of meteorite-sized pieces entering on a daily basis.

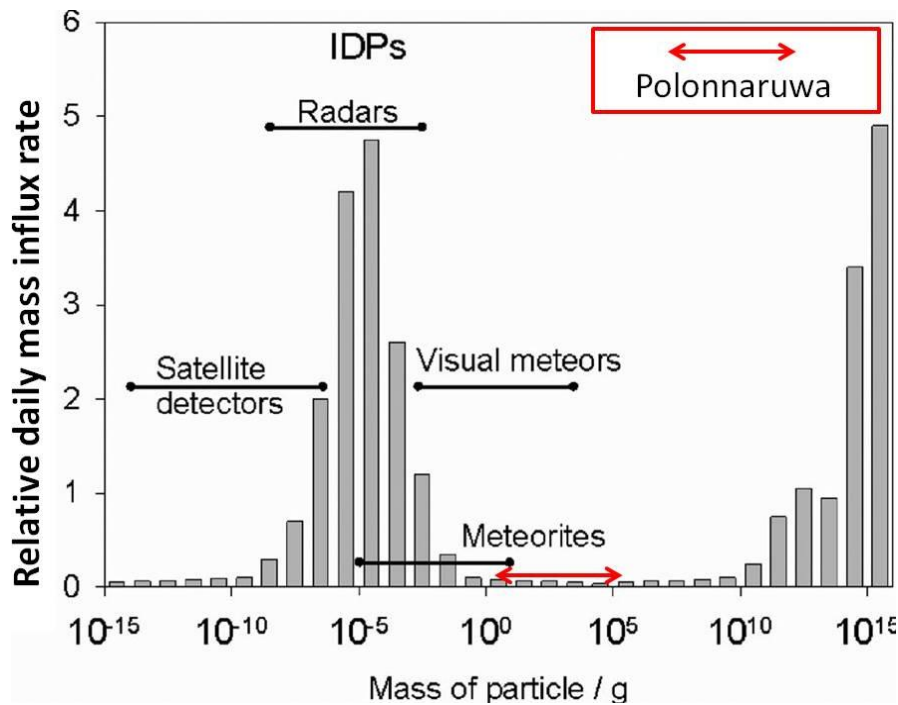

Fig.2 Relative mass influx of meteorites. The original plot from Flynn (2002) has to be scaled to accord with the revised values from Plane (2012). Polonnaruwa meteorite fragments are at the end of the Meteorites range.

**Collected Meteorites**

Estimating the planet-wide rate of meteorite entry from visual sightings of fireballs in the lower atmosphere is a complex matter depending on many imponderables, including the population density where a fall occurred, variable levels of public awareness and reporting, and the time of day/night, among other factors. It is to be expected therefore that cometary bolides that are seen to fall and come to be collected represents only a minute subset of what is actually incident on the Earth. The subset of meteorites thus collected after witnessed falls may be grouped in several broad categories:

- Stony meteorites, including chondrites, and carbonaceous chondrites;
- Stony-iron meteorites and iron meteorites.

These two broad classes of meteoroid of asteroidal or cometary origin have an overall density range of ~ 2 – 4.7 g cm$^{-3}$. Any meteorite with a significantly lower density will not readily be recognised as such, nor even tested or classified as a meteorite. According to such an arbitrary criterion the Polonnaruwa meteorite, that we have discussed in earlier papers (Wickramasinghe et al, 2013a,b), and which has a density of ~0.6 g cm$^{-3}$ will have gone unnoticed.

**Micrometeorite densities – a clue to a class of unnoticed meteorites**

Micrometeorites have for long been estimated to possess densities that span a range extending to much lower values than are normally found in meteorites. Verniani (1969) used a simple fragmentation model to estimate densities in the range 0.14 – 0.63 g cm$^{-3}$ for the majority of meteor showers, whereas Ceplecha (1977) obtained densities for the class of cometary meteoroids in the range 0.2-1 g cm$^{-3}$, using more realistic fragmentation models for a porous bolide. The most recent modelling of an extensive set of meteor data by Kikway et al (2011) yields the result displayed in Fig. 3 showing a distinction between three broad classes of incoming bolide:

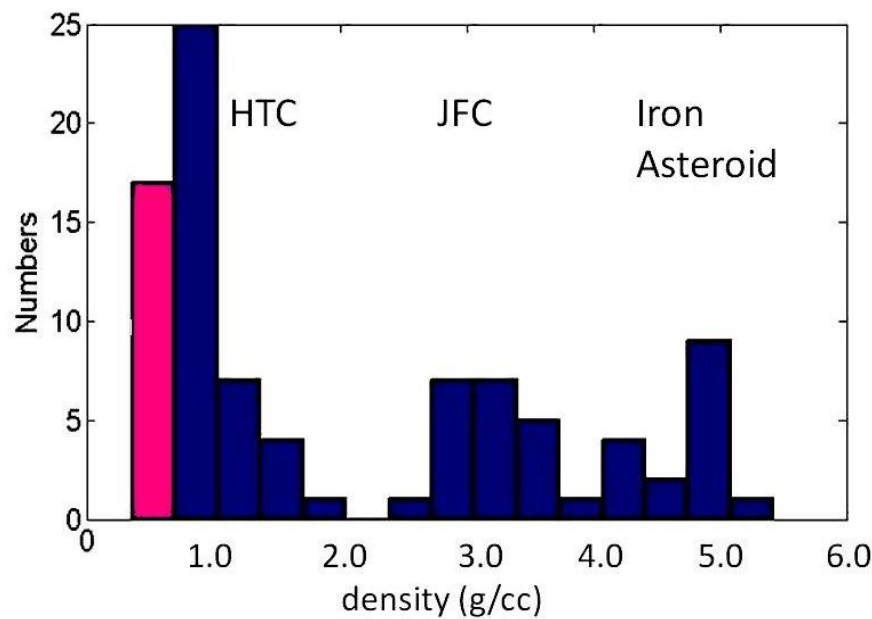

Fig.3 Distribution of average density calculated by Kikway et al (2011).

Cometary meteor (Halley-type comets, HTC)  density ~ 0.4-1 g cm$^{-3}$

Chondritic meteor (Jupiter family comet/asteroid, JFC)  density ~ 3 g cm$^{-3}$

Iron-rich meteor (asteroidal)  density ~5 g cm$^{-3}$

The pink bar in Fig. 3 shows the data for porous IDP's exhibiting a significant peak near the value of ~ 0.5 g cm$^{-3}$ for the average density. Integrating under the histogram to obtain cumulative mass fluxes, we find that the most porous particles contribute about 1/10 of the total mass. If we adopt 50 tonnes per day (lower limit) for the total micrometeorite input, we now get ~ 5 tonne per day in the form of porous micrometeorite particles on this basis.

Interplanetary dust particles of cometary origin (IDP's) have been recovered several years ago by D.E. Brownlee from heights of ~20 km using high-flying aircraft (Brownlee, Tomandl, and Olszewski, 1977). Fig. 4 shows such a Brownlee particle of chondritic composition. Independently, Harris et al (2002) have isolated large numbers of porous, fluffly particles of presumed cometary origin from sampling the stratosphere at heights of ~41 km using cryosamplers flown on balloons from Hyderabad, India. Fig. 5 shows an example of cometary micrometeoroid clump that was found in this way (Miyake, Wallis and Wickramasinghe, 2009. Miyake, et al, 2010). It is interesting to note that we have in Fig. 5 evidence for a silica rod that is strikingly similar to the diatoms that were found in the Polonnaruwa meteorite (Wickramasinghe, 2013a). Our estimate of the bulk density of the structures in Fig.4 and Fig.5 is ~0.6-0.8 g cm$^{-3}$, very similar to the value appropriate to the Polonnaruwa meteorites, and close to the peak in the micrometeorite data in Fig.3.

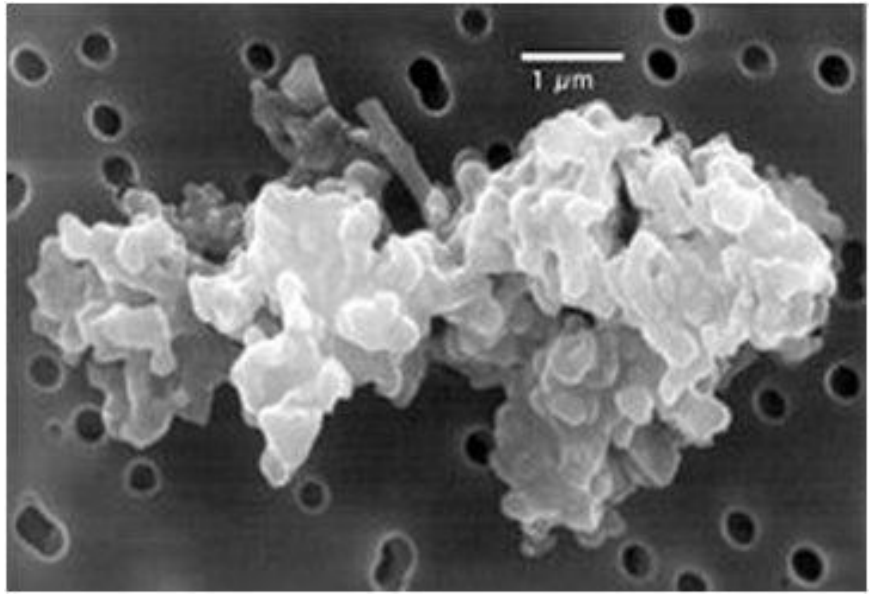

Fig. 4  Brownlee particle with carbonaceous chondritic composition

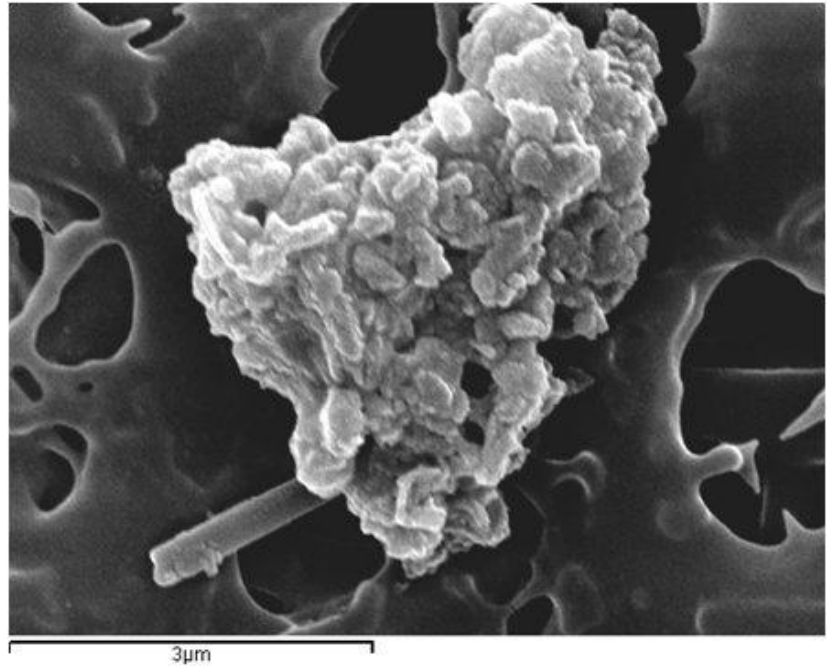

Fig. 5.  Cryoprobe sample from 41 km (Wallis *et al*, 2010)

These would mostly be of radii < 50μm, and for entry speeds in excess of ~ 20 km/s (see Table 1) would be mostly heated on entry to burn up as meteors.  This latter class of low density micrometeorites would inevitably be associated with larger aggregates of the same material and constitute a separate category of meteorite that survives entry into the lower stratosphere to disintegrate in the form of visible fireballs.

If our estimate of 0.5 tonnes of porous micrometeorites is to match an equal mass of unfragmented 1 metre bolides of the same composition, we would expect one such object exploding and disintegrating in the lower stratosphere every second day, on the average, to produce a Polonnaruwa-type event.  It is therefore reasonable to conclude that the Polonnaruwa meteorites represent larger composites of the type of structure that are found in miniature in Figs. 4 and 5.  The siliceous fibres in the latter can be matched to some of the diatom fossils found in the Polonnaruwa meteorite (Wickramasinghe et al, 2013a,b). This type of low density meteorite has gone largely unrecognised, as well as being under-counted in mass-histograms such as Fig.1.

**Acknowledgement**

We are grateful to Astrobiology Research Trust for their continued support of this work and to Nadika Udyakantha for assistance in connection with the collection of meteorite samples.